\documentclass{aastex}

\begin{document}

\title{Calibration of the Barnes-Evans relation using
interferometric observations of Cepheids}

\author{Tyler E. Nordgren\altaffilmark{1}, B. F. Lane\altaffilmark{2},
R. B. Hindsley\altaffilmark{3}, P. Kervella\altaffilmark{4}
}

\altaffiltext{1}{University of Redlands, Department of Physics,
1200 E Colton Ave, Redlands California}

\altaffiltext{2}{Department of Geological \& Planetary Sciences, MS 150-21,
California Institute of Technology, Pasadena CA 91125, U.S.A.}

\altaffiltext{3}{Remote Sensing Division, Naval Research Laboratory\\
Code 7210, Washington, DC 20375}

\altaffiltext{4}{European Southern Observatory, \\
Karl-Schwarzschildstr. 2, 85748 Garching, Germany}

\begin{abstract}

Direct diameter observations of Cepheid variables
are used to calibrate the Barnes-Evans Cepheid
surface brightness relation. Fifty-nine separate Cepheid diameter
measurements from four different optical interferometers
are used to calculate surface brightnesses as
a function of magnitude and color. The linear fit to
Cepheid surface brightness versus color is
in excellent agreement with functions
in the literature found using
interferometric observations of non-variable
giant and supergiant stars \citep{fag97}. Using these relations the
distance is calculated to $\delta$ Cephei, for which an
independent distance is known from trigonometric parallax.
The distance from the relation in this paper differs from that
derived by the previously published relation by
4\% but is still marginally within the
combined errors. Both distances are well within the
errors of the distance derived
from trigonometric parallax.

\end{abstract}


\section{Introduction}

The distance to the Large Magellanic Cloud (LMC) is of great
interest as it is the primary means of calibrating
extra-galactic distance scales.
One method for determining this distance is to determine
the distance to Cepheids within the LMC. The pulsation parallax method,
by which angular diameter variations
are coupled with linear diameter displacements, is one means by
which Cepheid distances are determined.

Whereas radial velocity
measurements are used to derive linear diameter changes,
angular diameters have, until recently,
only been estimated from photometric surface brightness relations \citep{wes69}.
Recent results from optical
interferometry have yielded mean angular diameters for the Cepheids
$\delta$ Cephei, $\eta$ Aquilae, $\zeta$ Geminorum, and Polaris
\citep{mou97,ten99, ten00}. Only
$\zeta$ Geminorum, however, has had its diameter variation measured directly
\citep{lan00}.
Thus, while direct interferometric observations are
confined to nearby Galactic Cepheids these observations can be used to
calibrate the surface brightness relations necessary for
estimating the angular diameters of more distant Cepheids, such
as those in the LMC.

\subsection{Surface Brightness Relations}

Several calibrations of Cepheid surface brightness relations 
have been used to calculate angular diameters of Cepheids \citep{wes69,
bae76, cor81, wel94, las95}. The Barnes-Evans
relation \citep{bae76} has both optical and infrared versions
\citep{wel94,fag97}. \citet{mab87} have
used these relations to
calculate the angular diameters of over a hundred Cepheids in the Galaxy and the Magellanic Clouds \citep{gmb99,gie00}.
In general these methods relate the surface brightness at a given
pulsation phase, $F$, in
some dereddened magnitude, $V_o$ or $K_o$, to the angular diameter,
$\theta$, at that phase:

\begin{equation}
F_{V_o} = 4.2207 -0.1V_o - 0.5\log\theta,
\end{equation}
\begin{equation}
F_{K_o} = 4.2207 -0.1K_o - 0.5\log\theta,
\end{equation}
\noindent where units for the surface brightness are Watts per meter squared per
micron. A relation (e.g., linear) is then sought between
the surface brightness as
a function of phase to the Cepheid color at that phase:

\begin{equation}
F_{V_o} = a + b(V-R)_o,
\end{equation}
\begin{equation}
F_{K_o} = c + d(J-K)_o.
\end{equation}

These versions
of the Barnes-Evans relation were most recently calibrated using
interferometric angular diameter observations of non-variable late-type
giant and supergiants stars \citep{fag97}.
In that work
Fouqu\'{e} and Gieren use the non-variable stellar observations to set
the zero-point of the relation between surface brightness (F$_{V_o}$,
F$_{K_o}$) and three colors: $(V-R)_o$, $(V-K)_o$, $(J-K)_o$. The slope for this
relation is derived directly from the Cepheid's magnitude and
linear displacement.
Using these slopes and the non-variable stars to supply a zero-point,
the full relation between angular diameter, magnitude and color as
a function of pulsation phase is found for Cepheid variables.
In this paper we use recent interferometric angular diameters
of three nearby Cepheids to fully calibrate the Barnes-Evans
surface brightness relation. 

Section 2 presents the interferometric data. Section 3 explores 
the agreement between observations of non-variable stars
by different interferometers. Section 4 presents the photometric
data analysis, including dereddening. Section 5 follows the analysis
of \citet{fag97} in order to derive the surface brightness relations.
Section 6 repeats this analysis for the single Cepheid, $\zeta$ Gem, for
which the most accurate angular diameters are available. Section 7
uses these relations to calculate a distance to the nearby Cepheid,
$\delta$ Cephei. Section 8 explores the limitations of the 
current analysis and the role of future, higher-precision
interferometric observations of Cepheids.

\section{The Data}

Interferometric observations have been acquired for three Cepheid
variables: $\delta$ Cephei, $\eta$ Aquilae, and $\zeta$ Geminorum.
These observations have taken place using the optical and
infrared interferometers: Grand Interf\'{e}rom\`{e}tre \`{a} 2 T\'{e}lescopes
(GI2T), Navy Prototype Optical Interferometer (NPOI),
Palomar Testbed Interferometer (PTI), and Infrared-Optical Telescope Array
(IOTA). A total of
59 separate observations of these three Cepheids have been reported in the
last four years \citep{mou97,ten00,lan00,ker01,arm01}.
In each case the reported diameter is a uniform-disk diameter
($\theta_U$) which assumes the star is a uniformly-bright disk. The
diameter used in Equations 1 and 2 is the more physically real limb-darkened
diameter ($\theta_L$). At the level of precision of these interferometers,
as given by the uncertainty in each uniform-disk diameter ($\sigma_{\theta}$),
a simple multiplicative limb-darkened conversion factor (LDC) is used
to convert uniform-disks to
limb-darkened disks. This factor is dependent upon the wavelength of
observation and the star's estimated effective temperature and surface
gravity (see \citet{lan00} for a representative derivation for the
LDC appropriate for PTI observations of $\zeta$ Gem).
Table 1 lists the uniform-disk angular diameters, their uncertainties and
the associated limb-darkened conversion factors. The Julian date and
derived pulsation phase for each observation (where phase 0 = maximum
light in V)
is also given.

\section{Nonvariable Giant Star Surface Brightness Relations}

For a sample of ten giant stars observed with the IOTA interferometer \citet{fag97} fit relations
between surface brightness in
$V_o$ and $K_o$ magnitude bands: F$_{V_o}$ and F$_{K_o}$, and color: $(V-R)_o$,
$(V-K)_o$, and $(J-K)_o$. Since 1997 many more 
stellar angular diameters have been published in the literature.
Does the increase
in the number of non-variable giants lead to a difference
in these relations (thereby changing their Cepheid
surface brightness calibration)? To answer this question
we use a sample of 57 non-variable giant
stars observed at the NPOI \citep{ten99,ten01} to calculate
the same relations between
F$_{V_o}$, F$_{K_o}$ and $(V-R)_o$, $(V-K)_o$, and $(J-K)_o$. 
\citet{fag97} used an unweighted linear least squares fit, while for
this study we use a linear least squares fit weighted by the
limb-darkened angular uncertainty. The reason for this weighting
is that the angular diameters in this study span a larger range
than in \citet{fag97} and therefore there is a larger range
of diameter uncertainties. The formal uncertainty in our
least squares fit is multiplied by the square-root of the reduced
chi-square.

For the 57 NPOI giants we find:

\begin{equation}
F_{V_o} = 3.925(\pm0.009) - 0.365(\pm0.009) (V-R)_o, N=57, rms=0.016,
\end{equation}
\begin{equation}
F_{V_o} = 3.934(\pm0.005) - 0.123(\pm0.002) (V-K)_o, N=57, rms=0.011,
\end{equation}
\begin{equation}
F_{K_o} = 3.942(\pm0.006) - 0.095(\pm0.007) (J-K)_o, N=57, rms=0.011.
\end{equation}
Compared to Fouqu\'{e}'s and Gieren's Equations (4) - (6):

$$F_{V_o} = 3.925(\pm0.017) - 0.379(\pm0.016) (V-R)_o, N=10, rms=0.012,$$
$$F_{V_o} = 3.930(\pm0.012) - 0.124(\pm0.004) (V-K)_o, N=10, rms=0.008,$$
$$F_{K_o} = 3.940(\pm0.013) - 0.100(\pm0.016) (J-K)_o, N=10, rms=0.008.$$

For all three cases the new relations are in excellent agreement
with the Fouqu\'{e} and Gieren relations. Interestingly, even though
the number of stars is almost six times greater, the root-mean-squares are
nearly equal and indicate an intrinsic width to the distribution.
\citet{moz01} find a similar scatter which they attribute to
a real spread within the class of stars.
Both this study and the
earlier one therefore start from the same
relationships between surface brightness (and hence interferometrically
measured angular diameter) and color.

As a representative example, Figure 1 displays F$_{V_o}$ versus $(V-K)_o$ for the
NPOI sample. The weighted linear least squares fit is given by the
solid line while the corresponding non-variable giants relation
of \citet{fag97}
is shown by the dashed line.

\section{The Cepheid Photometric Data}

For each Cepheid observation in the literature, \citep{mou97,ten00,ker01,lan00,arm01}
the phase is calculated for each diameter measurement using appropriate ephemerides.
In order to determine photometry at each of these phases,
spline fits are made to
the Cepheid photometric data in the literature. 

Stellar photometry comes from four sources.  Visual photometry is 
primarily from \citet{mab84}, with additional data for $\zeta$~Gem
from \citet{sho92}.  Infrared photometry for $\delta$ Cephei and $\eta$ Aql 
is from \citet{bar97}; infrared 
photometry for $\zeta$ Gem comes from \citet{waj68}.  The phases
given in \citet{mab84} have since been corrected to match the 
ephemerides in \citet{mab85}, which are consistent with those
of \citet{bar97}.  Overlapping 
visual photometry has been used to ensure that the phasing is correct.

The photometry are somewhat irregular in phase coverage.  To permit 
interpolation to values at particular phases, smooth curves were drawn by 
eye through the raw data, and cubic splines were then fit to these curves.
These fit curves result
in BVRIJK data at 0.01 phase intervals for each Cepheid.

The photometric curves are dereddened
using the standard equations.
The (B-V) color excess, E(B-V), for each Cepheid is taken from
\citet{fer90}. We adopt R$_V$ = 3.26 from \citet{car89}, and

\begin{equation}
E(V - R) = 0.97E(B - V)
\end{equation}
\begin{equation}
E(V - K) = 2.88E(B - V)
\end{equation}
\begin{equation}
E(J - K) = 0.56E(B - V)
\end{equation}

\noindent from \citet{fag97}, to
remain consistent with their analysis. This insistence upon
consistency with the earlier work is not strictly necessary as
that analysis shows the exact reddening law has an affect of less than 1\%
on the derived distance \citep{gie00}. This level of accuracy in
the determined distance is
currently smaller than the precision of the method by a factor
of two to three.

\section{Cepheid Surface Brightness Relations}

In the same manner as for the 57 non-variable giant stars,
a weighted least squares fit is made to the surface brightness
as a function of color for each of the 59 Cepheid observations.
The three relations found in this manner are:

\begin{equation}
F_{V_o} = 3.939(\pm0.006) - 0.364(\pm0.011) (V-R)_o, N=59, rms=0.026, \chi^2=1.90
\end{equation}
\begin{equation}
F_{V_o} = 3.956(\pm0.011) - 0.134(\pm0.005) (V-K)_o, N=59, rms=0.026, \chi^2=1.81
\end{equation}
\begin{equation}
F_{K_o} = 3.934(\pm0.009) - 0.080(\pm0.021) (J-K)_o, N=59, rms=0.026, \chi^2=2.25
\end{equation}

Figures 2 - 4 display F$_{V_o}$ versus $(V-R)_o$, F$_{V_o}$ versus $(V-K)_o$,
and F$_{K_o}$ versus $(J-K)_o$ respectively for the
59 Cepheid observations. The one Cepheid outlier in each figure
is a single observation of $\zeta$ Gem from IOTA (solid circle). At
IOTA's baseline and observational wavelength $\zeta$ Gem is only marginally
resolved, hence the poor precision of the observation (as evidenced by
the larger error bar). The solid lines are the weighted linear
least squares fits to the Cepheid observations. The dashed lines
are the corresponding relations from \citet{fag97} given in
their Table 3. The open circles are the non-variable giant
stars from the previous section (and Figure 1) which are shown for comparison.

For a star of spectral type A0 where $(V-R)_o$ = $(V-K)_o$ = $(J-K)_o$ = 0,
each of the relations should yield a common surface brightness.
While they are very similar, within the errors they are marginally unequal.
The weighted average of the intercepts for $(V-R)_o$, $(V-K)_o$, and
$(J-K)_o$ is 3.941 $\pm$ 0.004. Forcing this value to be the zero-point
for each of the three relations while solving for new slopes
yields the final relations:

\begin{equation}
F_{V_o} = 3.941(\pm0.004) - 0.368(\pm0.007) (V-R)_o,
\end{equation}
\begin{equation}
F_{V_o} = 3.941(\pm0.004) - 0.125(\pm0.003) (V-K)_o,
\end{equation}
\begin{equation}
F_{K_o} = 3.941(\pm0.004) - 0.096(\pm0.010) (J-K)_o,
\end{equation}

Using the non-variable stars to arrive at a Cepheid zero-point
(while also requiring there to be a common zero-point)
Fouqu\'{e} and Gieren derive for Cepheids (their Equations (26) - (28):

$$F_{V_o} = 3.947(\pm0.003) - 0.380(\pm0.003) (V-R)_o,$$
$$F_{V_o} = 3.947(\pm0.003) - 0.131(\pm0.003) (V-K)_o,$$
$$F_{K_o} = 3.947(\pm0.003) - 0.110(\pm0.003) (J-K)_o.$$

The two sets of
relations are consistent within the combined uncertainties at the 2$\sigma$
level (with the $(V-K_)o$ relation being the closest in agreement).
Figure 5 displays F$_{V_o}$ versus $(V-K)_o$
for the 59 Cepheid observations, where the solid line is the weighted linear
least squares fit to the Cepheid observations with a common
zero-point. The dashed line
is the corresponding relation from \citet{fag97} given in
their Equation (27).

\section{$\zeta$ Geminorum}

The vast majority of Cepheid observations do not detect
the pulsation of the Cepheid. Notice the scatter in points about the
best fit lines in Figure 5. It is possible that the lack of
precision of the non-PTI $\zeta$ Gem observations also
hides a lack of accuracy which could skew the relations found
in the previous sections. We therefore investigate
whether surface brightness relations fit to the PTI observations
alone reveal a statistically different result. In addition, since
\citet{gie88} report that there is
a weak period dependence on the slope for Cepheids, fitting a relation
to a single Cepheid will allow for the investigation of this dependence
within the data.  For the color $(V-R)_o$, this dependence is of the form

\begin{equation}
m = -0.359 - 0.020\log P
\end{equation}

\noindent which for $\zeta$ Gem with P = 10.1507 days, yields $m = -0.379$.

Using only the highest precision PTI data to recalibrate the
surface brightness relations yields (where we have not constrained
the relations to a common zero-point):

\begin{equation}
F_{V_o} = 3.940(\pm0.013) - 0.366(\pm0.024) (V-R)_o,
\end{equation}
\begin{equation}
F_{V_o} = 3.985(\pm0.014) - 0.152(\pm0.008) (V-K)_o,
\end{equation}
\begin{equation}
F_{K_o} = 3.980(\pm0.030) - 0.180(\pm0.050) (J-K)_o,
\end{equation}

The slope for the $(V-R)_o$ relation from the PTI $\zeta$ Gem data is
consistent with the P = 10.1507 day slope of $m = -0.379$.
It also agrees well with Equation 11 derived from the entire
sample of Cepheids. The two IR relations (Equations 19 and 20) are marginally
inconsistent with their counterparts from the entire Cepheid sample.
One possible reason for the difficulty with the IR relations
is the lack of good J and K magnitudes for $\zeta$ Gem. The only
published J and K magnitudes for this Cepheid are by \citet{waj68}
in the Communications of the Lunar and Planetary Lab.
Figure 6 shows F$_{V_o}$ versus $(V-K)_o$
for the PTI $\zeta$ Gem observations, where the solid line is the weighted linear
least squares fit to the Cepheid observations from Equation 19.
The dashed line
is the corresponding fit to all Cepheids from Equation 15.

%

Using Equations 14 - 16 the new
Cepheid angular diameter relations are:

\begin{equation}
\log\theta = 0.5594 - 0.2 V_o + 0.736 (V-R)_o
\end{equation}
\begin{equation} 
\log\theta = 0.5594 - 0.2 V_o + 0.250 (V-K)_o
\end{equation}
\begin{equation} 
\log\theta = 0.5594 - 0.2 K_o + 0.192 (J-K)_o
\end{equation}

\section{Distance to $\delta$ Cephei}

With angular diameters from photometry and linear displacements from
radial velocities the distance and mean linear radius of a Cepheid can be calculated.
We have chosen to calculate the distance to the nearby Cepheid, $\delta$ Cephei,
for which an independent distance is known from trigonometric parallax
\citep{ten00}.

$F_{V_o}$ is calculated using
50 V magnitudes from \citet{mab84}. $F_{K_o}$ is calculated using the
same 50 V magnitudes plus the interpolated (V-K) color curve
which yields 50 values for the K magnitude.
These magnitudes plus the interpolated
color curves from the previous section yield the angular diameter
of $\delta$ Cephei as a function of pulsation phase.

Radial velocity data for $\delta$ Cephei are taken from \citet{but93} and 
\citet{sha58}.  As described by \citet{but93}, the data of \citet{sha58} 
were retarded in phase by 0.117, which corrected the data to the ephemerides
of \citet{mab85}.  The Shane data were also increased by a constant
2.379 km/s to match the \citet{but93} values.  As with the photometry, smooth
curves were drawn through the data and fitted with cubic splines in order to 
interpolate in phase.

Radial velocities are first converted to pulsation velocities using the
projection factor for $\delta$ Cephei, p = 1.31 $\pm$ 0.03 \citep{par72}.
It is important to note that the projection factor has a direct
effect on the final distance determination. For a Cepheid of similar period
to $\delta$ Cephei (CV Mon with period = 5.378 days) \citet{gfg97}
use a factor of 1.368, differing from this paper by 4.4\%.
This difference leads
directly to a 4.4\% difference in the final distance. However, since this
paper is primarily concerned with comparing different surface brightness relation
calibrations, the comparison is valid provided the same p-factor is used in
both cases (which has been done).

Linear radii displacements are next found by integrating the radial velocities.
These displacements are then matched to the angular diameters found using
the photometry.  These curves were fitted to each other iteratively, because the 
angular diameters depend weakly on the assumed surface gravity, which in turn depends
on the assumed mean linear diameter and mean surface gravity.  Level effects in stellar
atmospheres would admit a slight phase lag in the linear radii displacements, relative
to the angular diameters. In fact the best fit was found with zero phase shift.

Using these angular diameters and linear displacements the distance and mean radius
for $\delta$ Cephei is found from the equation:

\begin{equation}
D_o + \Delta D(t) = 10^{-3} d \times \theta(t)
\end{equation}

\noindent where $D_o$ is the mean linear diameter in AU,
$\Delta D(t)$ is the linear displacement in AU at a time $t$,
$d$ is the distance in parsecs and $\theta(t)$ is the angular
diameter in milliarcseconds at time $t$. Figures 7 and 8 display
angular diameter versus linear displacement and angular diameter as
a function of phase for $\delta$ Cephei from the $(V-K)_o$ relation in Equation 22.

Table 2 column 2 lists the newely calculated distances to $\delta$ Cephei
using angular diameters from Equations 21 - 23. The weighted mean of the three relations
is 272 $\pm$ 6 pc. Using the three equations from \citet{fag97}
the same data yield the distances in column 3. The weighted mean
of the three distances in column 3 is 262 $\pm$ 5 pc. All but the $(V-K)_o$ distance
are consistent between the two sets of equations (and that relation is
only just outside the mutual errors). The mean distances differ by 4\%
and are marginally
consistent at the one-sigma level.
The triginometric parallax distance of $\delta$ Cephei as calculated from
the Hipparcos parallax and USNO parallax is
278$^{+48}_{-36}$ pc \citep{ten00} which is consistent with both.

As noted earlier, had we used a higher value of p-factor as done by \citet{gfg97}
the distances we calculated using both surface brightness relations would be
4.4\% smaller. Both distances would still be well within the Hipparcos parallax
distance.

\section{Further Work and Limitations}

Over the next year the NPOI will increase its resolution by nearly
a factor of two which will add at least two more Cepheids to the
sample list and increase the precision of those in the current sample.
In the coming year
VLTI baselines of up to 202 meters will make it possible to study
the pulsation of nine Cepheids covering a wide
range of periods. The position of the VLTI in the
Southern hemisphere makes it a valuable complement to the NPOI and PTI
instruments. The result is that soon there should be many more observations
of the quality of the PTI data. Such observations will improve the
quality of this analysis. For instance, as in \citet{lan00} where accurate and precise 
observed angular diameters as a function of phase lead to a Cepheid's distance
directly, new observations of $\delta$ Cephei will yield an independent
check on the results of the current surface brightness calibration.
Other questions to be addressed include whether or not it is possible to detect
the period dependence of the slope as found by \citet{gie88}.

In addition, newer, better observations are also expected to reveal
the limitations of this current analysis. For instance, the linear
relation fit to the Cepheid data assumes that the atmosphere is in
thermal equilibrium at all stages of the pulsation. While this may
be an acceptable approximation for the relatively small, short period
Cepheids in this sample, it is almost certainly not the case for
longer period Cepheids with extended atmospheres. More precise
observations which include those of longer period Cepheids will
reveal whether the surface brightness as a function of color deviates
from the simple linear fits used here.

\section{Conclusions}
Direct angular diameter observations of Cepheid variables
are used to calibrate the Barnes-Evans Cepheid
surface brightness relation. The linear fit to
Cepheid surface brightness versus color is
in good agreement with expressions in the literature by
\citet{fag97} found using
interferometric observations of non-variable
giant and supergiant stars. Using these relations the
distance is calculated to $\delta$ Cephei.
The distance from the relation in this paper differs from that
derived by the previously published relation by only
4\% which is marginally within the
combined errors. As much of this disagreement is likely due to
the high degree of scatter in the interferometric
diameters, it is expected that as interferometers improve, the
differences will diminish. Currently, however, both distances are
well within the
errors of the distance derived directly
from trigonometric parallax.

\clearpage

\begin{deluxetable}{llcllllll}
\footnotesize
\tablecaption{Cepheid Angular Diameters\label{tbl-1}}
\tablewidth{0pt}
\tablehead{
\colhead{Cepheid} & \colhead{Obs.}  & \colhead{$\lambda_{obs}$} &
\colhead{LDC} &\colhead{JD} & \colhead{Phase} &
\colhead{$\theta_U$} & \colhead{$\sigma_{\theta}$} & \colhead{ref} \\ 
\colhead{} & \colhead{}  & \colhead{($\mu$m)} & \colhead{} &
\colhead{} & \colhead{} &
\colhead{(mas)} & \colhead{(mas)}
}
\startdata
$\delta$ Cephei & NPOI &0.740& 1.044 & 2450788.63 & 0.80 &   1.55 & 0.06 & a\\
                &      &     &       & 2450994.91 & 0.24 &   1.63 & 0.09 & \\
                &      &     &       & 2450995.93 & 0.42 &   1.42 & 0.11 & \\
                &      &     &       & 2450996.97 & 0.61 &   1.48 & 0.21 & \\
                &      &     &       & 2450997.93 & 0.80 &   1.36 & 0.11 & \\
                &      &     &       & 2450998.93 & 0.98 &   1.27 & 0.12 & \\
                &      &     &       & 2451007.96 & 0.66 &   1.52 & 0.10 & \\
                &      &     &       & 2451008.92 & 0.85 &   1.33 & 0.08 & \\
                &      &     &       & 2451009.96 & 0.03 &   1.48 & 0.07 & \\
                &      &     &       & 2451010.92 & 0.22 &   1.54 & 0.07 & \\
                &      &     &       & 2451011.91 & 0.40 &   1.47 & 0.07 & \\
                &      &     &       & 2451012.90 & 0.59 &   1.50 & 0.07 & \\
                &      &     &       & 2451088.81 & 0.71 &   1.32 & 0.12 & \\
                &      &     &       & 2451089.78 & 0.89 &   1.46 & 0.07 & \\
                &      &     &       & 2451093.76 & 0.64 &   1.41 & 0.02 & \\
                &      &     &       & 2451097.78 & 0.38 &   1.47 & 0.05 & \\
                &      &     &       & 2451098.85 & 0.57 &   1.62 & 0.06 & \\
		&&&&&&&&\\
		& GI2T & 0.669 - 0.675 & 1.045 & 2449566.6  & 0.05 &
1.31\tablenotemark{\dagger} & 0.59 & b \\
		&      &     &       & 2449572.5  & 0.17 &   1.46 & 0.73 & \\
		&      &     &       & 2449642.3  & 0.17 &   1.37 & 0.54 & \\
		&      &     &       & 2449643.3  & 0.36 &   1.97 & 0.41 & \\
		&      &     &       & 2449541.6  & 0.40 &   1.69 & 0.48 & \\
		&      &     &       & 2449569.5  & 0.61 &   1.58 & 0.48 & \\
		&      &     &       & 2449570.5  & 0.79 &   1.26 & 0.80 & \\
		&      &     &       & 2449640.3  & 0.80 &   1.71 & 0.44 & \\
		&      &     &       & 2449571.5  & 0.98 &   1.59 & 0.41 & \\
		&&&&&&&&\\
$\eta$ Aquilae  & NPOI & 0.740 & 1.048 & 2450638.86 & 0.98 &   1.87 & 0.08 & a\\
                &      &     &       & 2450640.88 & 0.29 &   1.70 & 0.09 \\
                &      &     &       & 2450641.86 & 0.39 &   1.85 & 0.08 \\
                &      &     &       & 2450997.83 & 0.00 &   1.62 & 0.19 \\
                &      &     &       & 2450998.88 & 0.17 &   1.70 & 0.06 \\
                &      &     &       & 2451007.88 & 0.43 &   1.83 & 0.06 \\
                &      &     &       & 2451008.91 & 0.57 &   1.53 & 0.20 \\
                &      &     &       & 2451009.85 & 0.67 &   1.69 & 0.08 \\
                &      &     &       & 2451010.84 & 0.81 &   1.39 & 0.08 \\
                &      &     &       & 2451011.84 & 0.95 &   1.44 & 0.07 \\
                &      &     &       & 2451012.87 & 0.89 &   1.53 & 0.10 \\
		&&&&&&&&\\
$\zeta$ Geminorum & NPOI & 0.740 & 1.046 & 2451098.98 & 0.52 &  1.49 & 0.21 & c\\
                &      &     &       & 2451229.83 & 0.41 &   1.39 & 0.22 \\
                &      &     &       & 2451232.72 & 0.70 &   1.54 & 0.05 \\
                &      &     &       & 2451233.71 & 0.80 &   1.44 & 0.06 \\
		&&&&&&&&\\
                & PTI  & 1.64& 1.038 & 2451605.73 & 0.45 & 1.679 & 0.014 & d\\
                &      &     &       & 2451606.74 & 0.55 &  1.678 & 0.046 \\
                &      &     &       & 2451607.78 & 0.65 &  1.651 & 0.057 \\
                &      &     &       & 2451614.69 & 0.33 &  1.800 & 0.059 \\
                &      &     &       & 2451615.68 & 0.43 &  1.740 & 0.031 \\
                &      &     &       & 2451617.67 & 0.63 &  1.590 & 0.028 \\
                &      &     &       & 2451618.64 & 0.72 &  1.538 & 0.007 \\
                &      &     &       & 2451619.67 & 0.82 &  1.553 & 0.018 \\
                &      &     &       & 2451620.67 & 0.92 &  1.589 & 0.028 \\
                &      &     &       & 2451622.7  & 0.12 &  1.677 & 0.046 \\
                &      &     &       & 2451643.66 & 0.19 &  1.666 & 0.012 \\
		&&&&&&&&\\
                & IOTA & 2.14& 1.02  & 2451259.78 & 0.35 & 2.040 & 0.29 & e \\
                &      &     &       & 2451262.72 & 0.64 & 1.602 & 0.39 \\
                &      &     &       & 2451262.76 & 0.64 & 1.826 & 0.43 \\  
                &      &     &       & 2451595.84 & 0.46 & 0.887 & 0.44 \\  
                &      &     &       & 2451595.87 & 0.46 & 1.737 & 0.58 \\  
                &      &     &       & 2451602.73 & 0.14 & 1.826 & 0.33 \\
                &      &     &       & 2451602.79 & 0.14 & 1.899 & 0.29 \\
\enddata
\tablenotetext{\dagger}{For the GI2T observations, $\theta_U$ is calculated
from the published visibilities, projected baseline, and mean
observational wavelength.}
\tablerefs{
(a) Armstrong et al. 2001; (b) Mourard et al. 1997; (c) Nordgren et al. 2001;
(d) Lane et al. 2000; (e) Kervella et al. 2001.}

\end{deluxetable}

\clearpage

\begin{deluxetable}{lllcc}
\footnotesize
\tablecaption{Distances to $\delta$ Cephei\label{tbl-2}}
\tablewidth{0pt}
\tablehead{
\colhead{Color} & \colhead{d}   & \colhead{d$_{FG}$\tablenotemark{a}} \\ 
\colhead{} & \colhead{(pc)}  & \colhead{(pc)}
}
\startdata
(V-R)  & 261 $\pm$ 17 & 257 $\pm$ 18 \\
(V-K)  & 269 $\pm$  7 & 254 $\pm$  6 \\ 
(J-K)  & 278 $\pm$  9 & 276 $\pm$  8 \\
{\bf Mean}   & 272 $\pm$  6 & 262 $\pm$  5 \\
\enddata

\tablenotetext{a}{Using the same photometry as Column 2 but the
surface brightness equations of \citet{fag97} }

\end{deluxetable}

\clearpage

\begin{figure}
\plotone{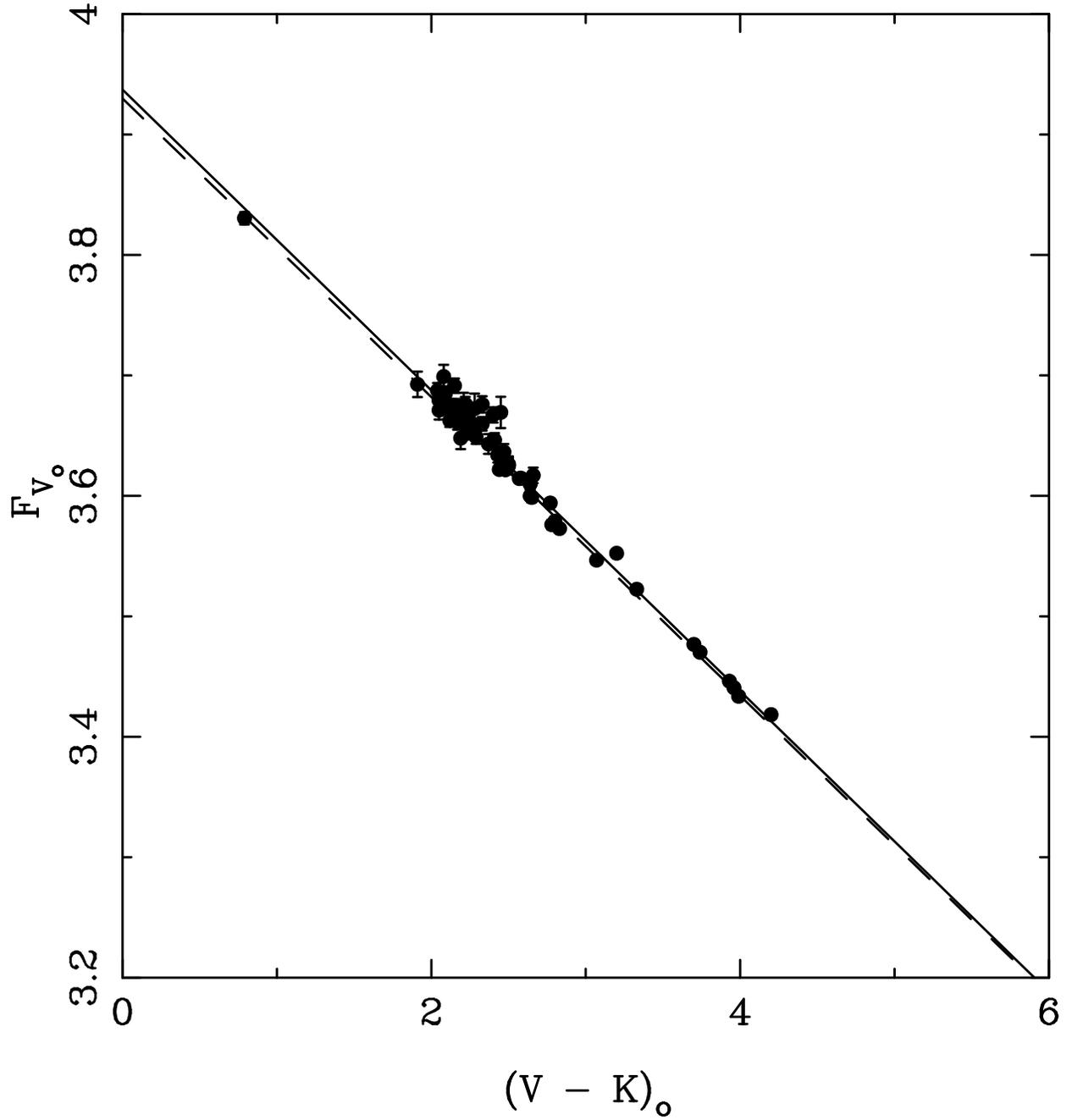}
\figcaption[nordgren.fig1.ps]{Dereddened F$_V$ versus (V-K) for 57 giant
stars observed with the NPOI. Solid line is the weighted least squares
fit. Dashed line is the least squares fit of \citet{fag97} for 10
giant stars observed with IOTA.}
\end{figure}

\clearpage

\begin{figure}
\plotone{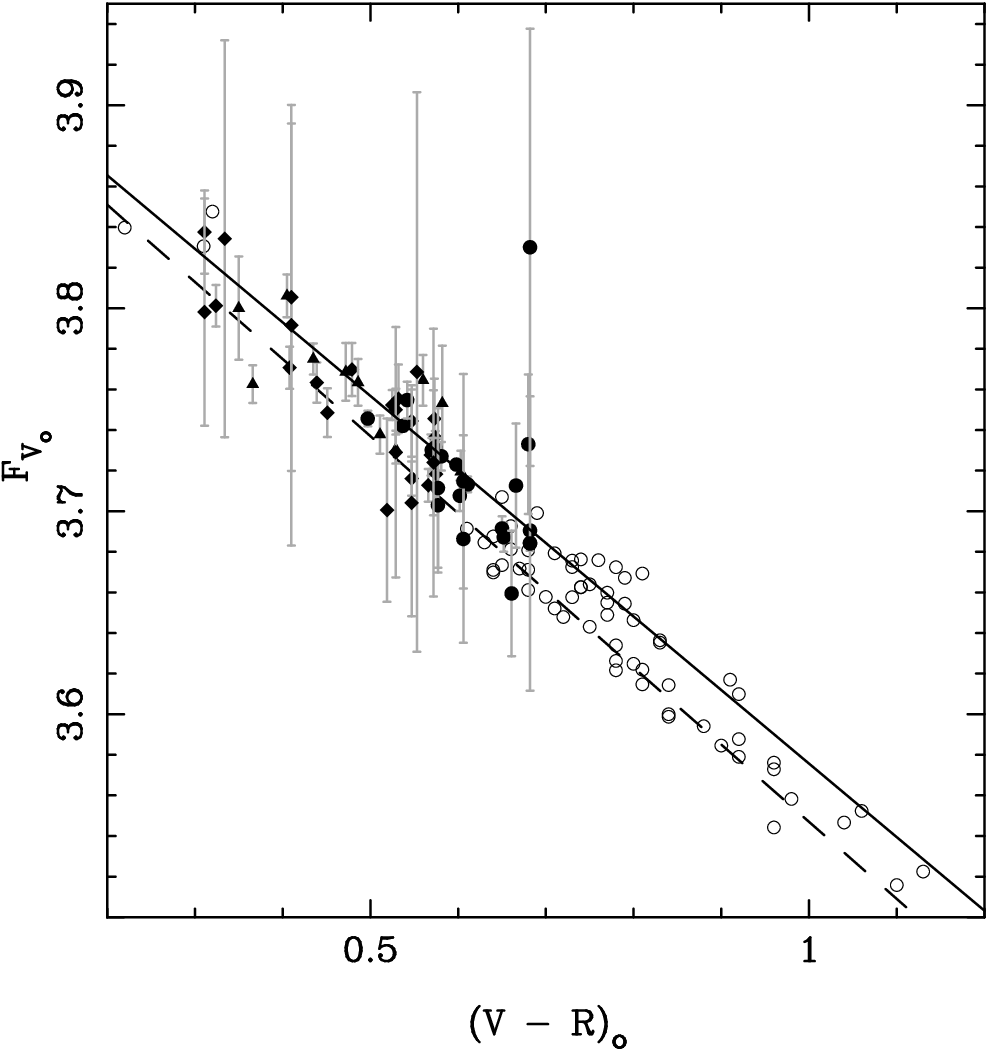}
\figcaption[nordgren.fig2.ps]{Dereddened F$_V$ versus (V-R) for 59 Cepheid
observations. Filled circles are $\zeta$ Gem, diamonds are $\delta$ Cep,
and triangles are $\eta$ Aql. Open circles are the non-variable giants.
The solid line is the weighted linear least squares
fit to just the Cepheid data. Dashed line is the linear least squares fit
of \citet{fag97} from their Table 3.}
\end{figure}

\clearpage

\begin{figure}
\plotone{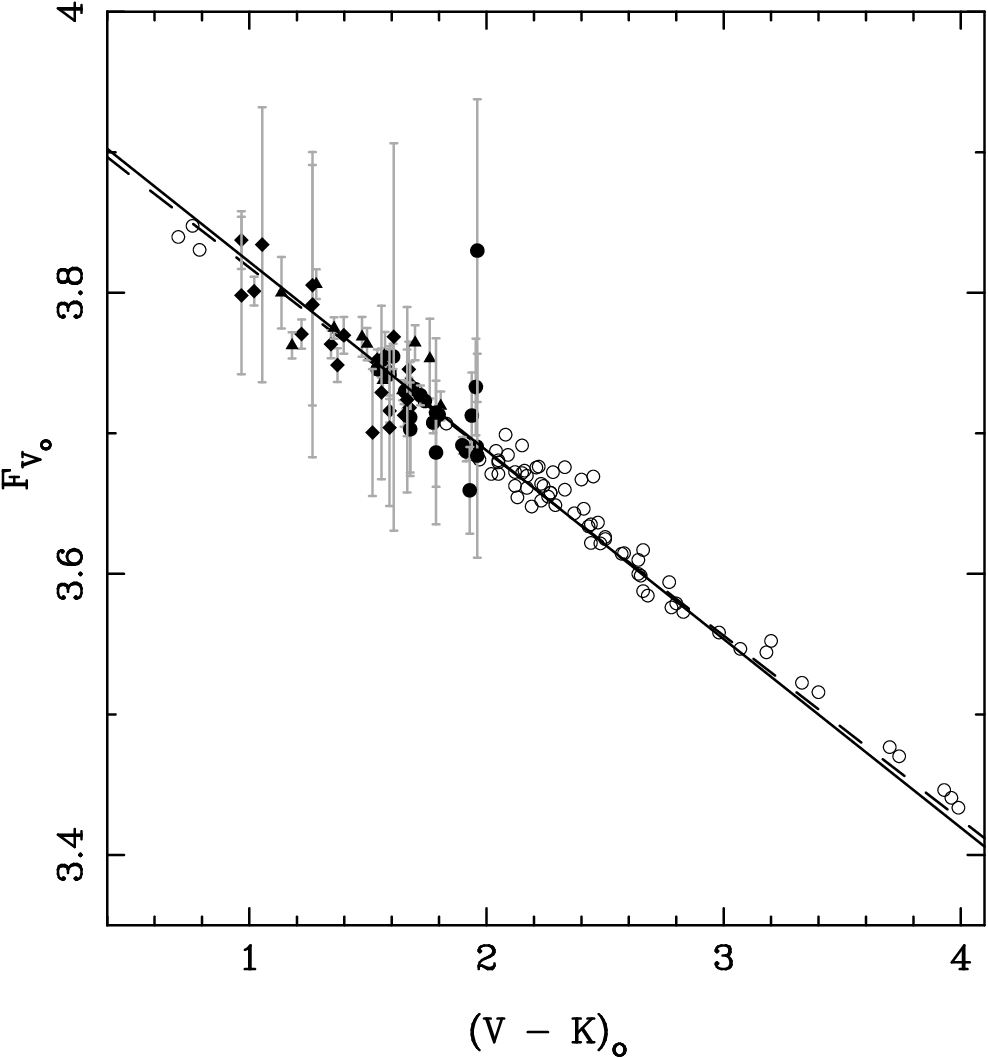}
\figcaption[nordgren.fig3.ps]{Dereddened F$_V$ versus (V-K) for 59 Cepheid
observations. Filled circles are $\zeta$ Gem, diamonds are $\delta$ Cep,
and triangles are $\eta$ Aql. Open circles are the non-variable giants.
The solid line is the weighted linear least squares
fit to just the Cepheid data. Dashed line is the linear least squares fit
of \citet{fag97} from their Table 3.}
\end{figure}

\clearpage

\begin{figure}
\plotone{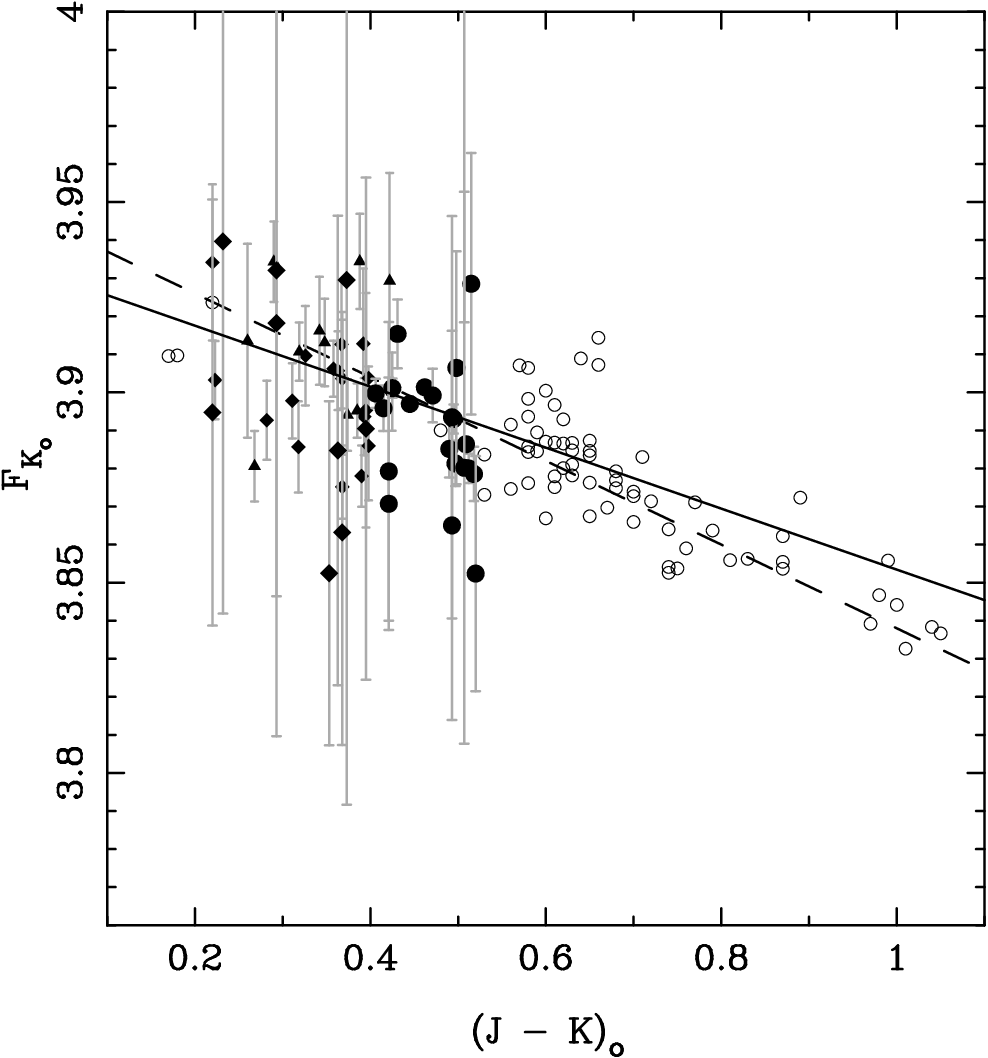}
\figcaption[nordgren.fig4.ps]{Dereddened F$_K$ versus (J-K) for 59 Cepheid
observations. Filled circles are $\zeta$ Gem, diamonds are $\delta$ Cep,
and triangles are $\eta$ Aql. Open circles are the non-variable giants.
The solid line is the weighted linear least squares
fit to just the Cepheid data. Dashed line is the linear least squares fit
of \citet{fag97} from their Table 3.}
\end{figure}

\clearpage

\begin{figure}
\plotone{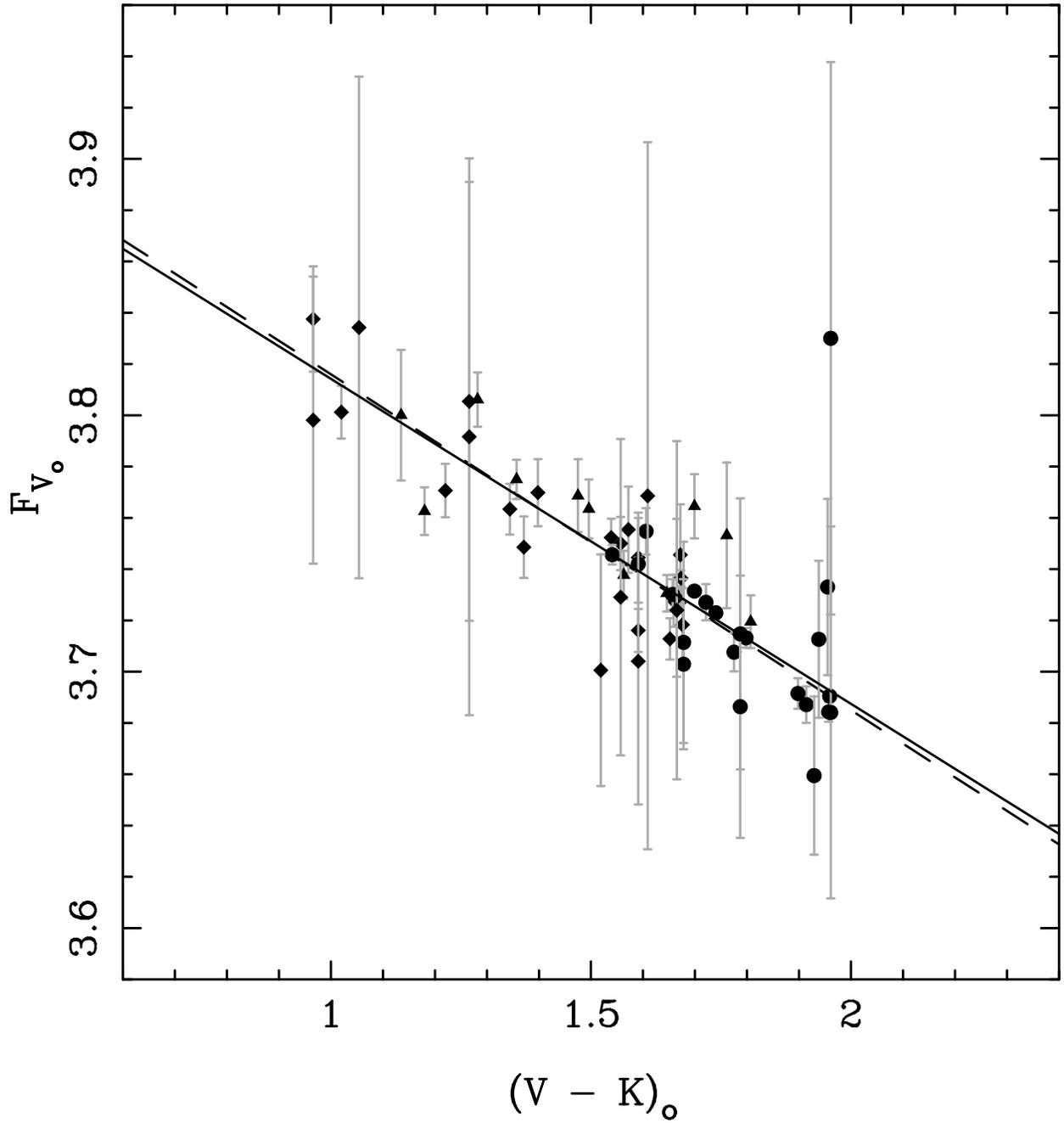}
\figcaption[nordgren.fig5.ps]{Dereddened F$_V$ versus (V-K) for 59 Cepheid
observations where the zero-point is constrained to be 3.941.
The solid line is the weighted linear least squares
fit. Dashed line is the linear least squares fit
of \citet{fag97} from their Equation (27).}
\end{figure}

\clearpage

\begin{figure}
\plotone{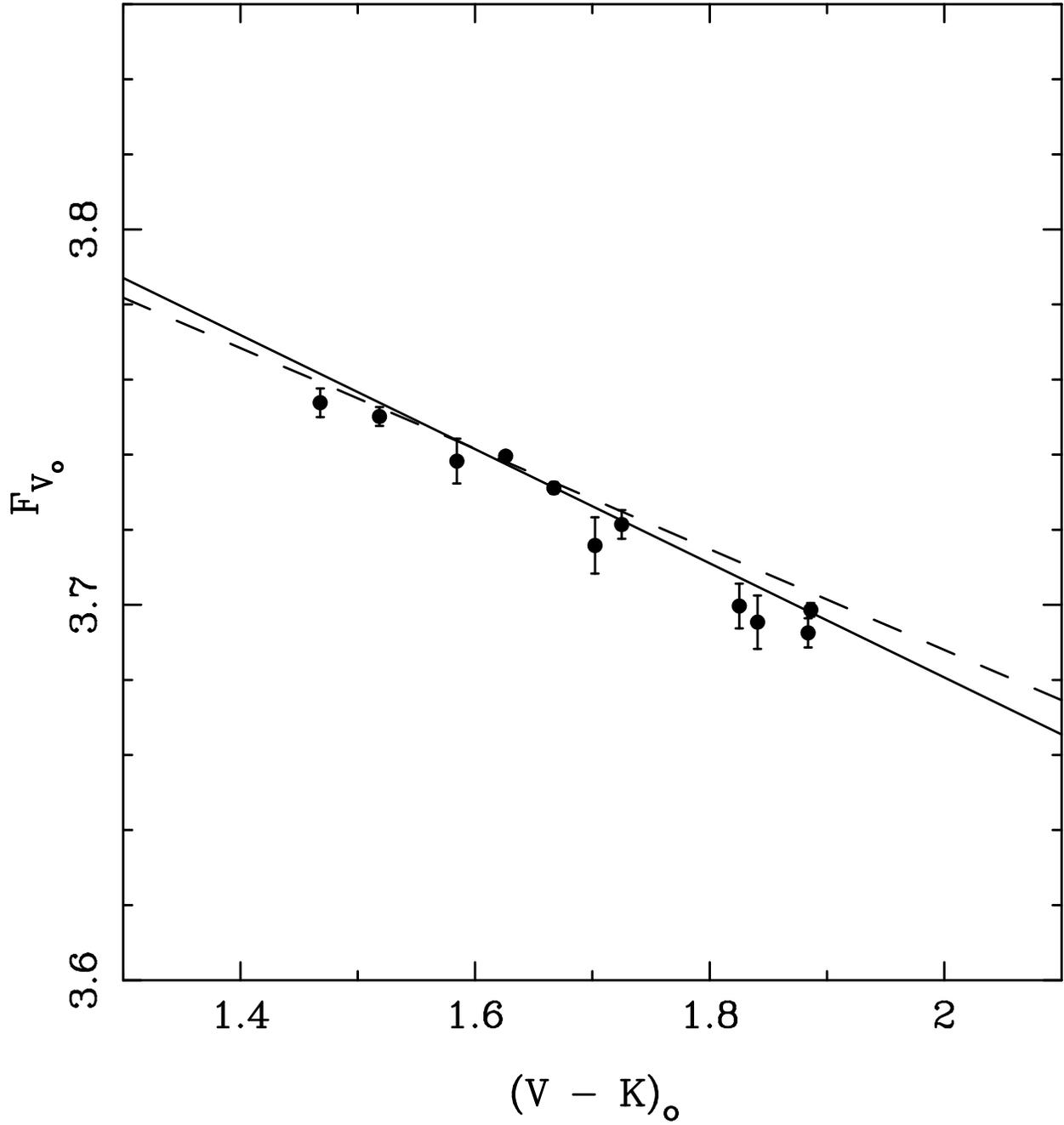}
\figcaption[nordgren.fig6.ps]{Dereddened F$_V$ versus (V-K) for the PTI 
$\zeta$ Gem data only. The solid line is the weighted linear least squares
fit. Dashed line is the weighted linear least squares fit
for all of the Cepheids.}
\end{figure}

\clearpage

\begin{figure}
\plotone{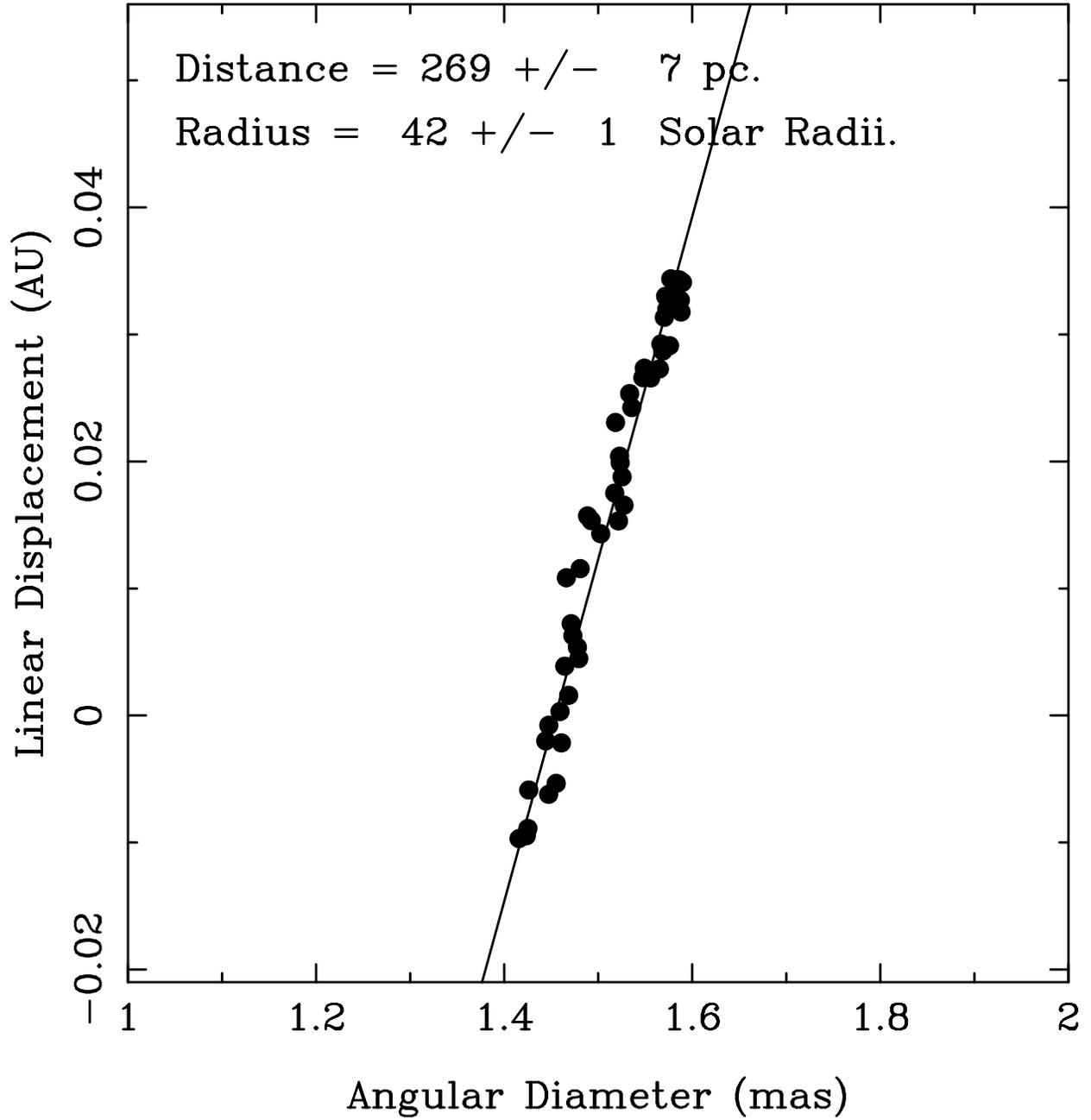}
\figcaption[nordgren.fig7.ps]{Calculated (V-K) angular diameter versus linear
displacement for $\delta$ Cep.}
\end{figure}

\clearpage

\begin{figure}
\plotone{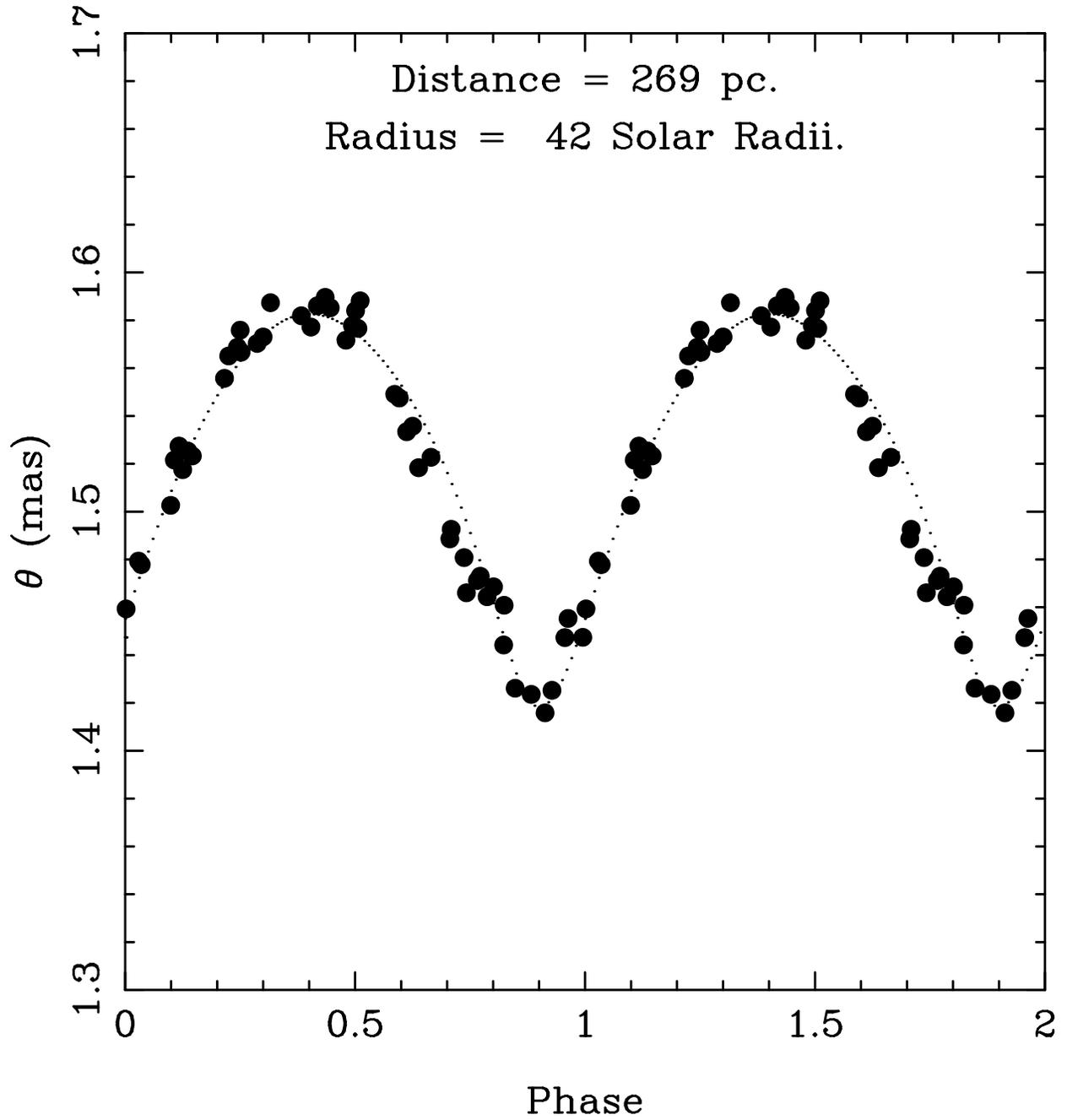}
\figcaption[nordgren.fig8.ps]{Calculated (V-K) angular diameter versus pulsation
phase for $\delta$ Cep. The dotted line is the best fit angular diameter
change given the observed linear diameter displacement and the derived
distance and mean diameter.}
\end{figure}


\begin{thebibliography}{}

\bibitem[Armstrong et al.(2001)]{arm01} Armstrong, J. T., Nordgren, T. E., Mozurkewich,
Germain, M. E., Hajian, A. R., Hindsley, R. B., Hummel, C. A., \& Thessin, R. N.
2001, \aj, 121, 476

\bibitem[Barnes \& Evans(1976)]{bae76}Barnes, T. G. \& Evans, D. S. 1976,
\mnras, 174, 489

\bibitem[Barnes et al.(1997)]{bar97}Barnes, T. G., Fernley, J. A., Frueh, M. L,
Navas, J. G., Moffett, T. J., \& Skillen, I. 1997, \pasp, 109, 645

\bibitem[Butler(1993)]{but93}Butler, R. P. 1993, \apj, 415, 323

\bibitem[Caccin et al.(1981)]{cor81}Caccin, B., Onnembo, A., Russo, G., \& Sollazzo, C. 1981, \aap, 97, 104

\bibitem[Cardelli, Clayton, \& Mathis(1989)]{car89}Cardelli, J. A., Clayton, G. C., \&
Mathis, J. S., 1989, \apj, 345, 245

\bibitem[Fernie(1990)]{fer90}Fernie, J. D., 1990, \apjs, 72, 153

\bibitem[Fouqu\'{e} \& Gieren(1997)]{fag97}Fouqu\'{e}, P. \& Gieren, W. P.
1997, \aap, 320, 799

\bibitem[Gieren(1988)]{gie88}Gieren, W. P. 1988, \apj, 329, 790



\bibitem[Gieren, Fouqu\'{e}, \& Gomez(1997)]{gfg97}Gieren, W. P., Fouqu\'{e}, P.,
\& Gomez, M. 1997, \apj, 488, 74

\bibitem[Gieren, Moffett, \& Barnes(1999)]{gmb99}Gieren, W. P., Moffett, T. J., \&
Barnes~III, T. G. 1999, \apj, 512, 553

\bibitem[Gieren et al.(2000)]{gie00}Gieren, W. P., Storm, J., Fouqu\'{e}, P.,
Mennickent, R. E., \& Gomez, M. 2000, \apjl, 533, L107

\bibitem[Kervella et al.(2001)]{ker01}Kervella, P., Coud\'{e} du Foresto, V., 
Perrin, G., Sch\"{o}ller, M., Traub, W. A., \& Lacasse, M. G. 2001, A\&A,
367, 876

\bibitem[Lane et al.(2000)]{lan00}Lane, B. F., Kuchner, M. J., Boden, A. F., Creech-Eakman, M., Kulkarni, S. R.
2000, Nature, 407, 485

\bibitem[Laney \& Stobie(1995)]{las95}Laney, C. D. \& Stobie, R. S.
1995, \mnras, 274, 337

\bibitem[Moffett \& Barnes(1984)]{mab84}Moffett, T. J., \&
Barnes~III, T. G. 1984, \apjs, 55, 389

\bibitem[Moffett \& Barnes(1985)]{mab85}
Moffett, T. J, \& Barnes~III, T. G. 1985, \apjs, 58, 843

\bibitem[Moffett \& Barnes(1987)]{mab87}Moffett, T. J., \&
Barnes~III, T. G. 1987, \apj, 323, 280

\bibitem[Mourard et al.(1997)]{mou97}Mourard, D., Bonneau, D., Koechlin, L., Labeyrie, A.,
Morand, F., Stee, P., Tallon-Bosc, I., \& Vakili, F. 1997, \aap, 317, 789

\bibitem[Mozurkewich et al.(2001)]{moz01}Mozurkewich, D., Elias II, N. M., 
Hajian, A. R., Johnston, K. J., \& Armstrong, J. T. 2001, \aj, submitted

\bibitem[Nordgren et al.(1999)]{ten99}Nordgren, T. E., Germain, M. E., 
Benson, J. A., Mozurkewich, D., Sudol, J. J., Elias II, N. M., 
Hajian, A. R., White, N. M., Hutter, D. J., Johnston, K. J.,
Gauss, F. S., Armstrong, J. T., Pauls, T. A., \& Rickard, L. J 1999, \aj,
118, 3032

\bibitem[Nordgren et al.(2000)]{ten00}Nordgren, T. E., Armstrong, J. T.,
Germain, M. E., Hindsley, R. B., 
Hajian, A. R., Sudol, J. J., \& Hummel, C. A. 2000, ApJ, 543, 972

\bibitem[Nordgren, Sudol, \& Mozurkewich(2001)]{ten01}Nordgren, T. E., Sudol, J. J.,
\& Mozurkewich, D. 2001, \aj, 122, 2707

\bibitem[Parsons(1972)]{par72}Parsons, S. B. 1972, \apj, 174, 57

\bibitem[Shane(1958)]{sha58}Shane, W. W. 1958, \apj, 127, 573

\bibitem[Shobbrook(1992)]{sho92}Shobbrook, R. R. 1992, \mnras, 255, 486


\bibitem[Welch(1994)]{wel94}Welch, D. L. 1994, \aj, 108, 1421

\bibitem[Wesselink(1969)]{wes69}Wesselink, A. J. 1969, \mnras, 144, 297

\bibitem[Wisniewski \& Johnson(1968)]{waj68}Wisniewski, W. Z. \& Johnson,
H. L. 1968, Comm. Lunar and Planet. Lab., 7, 57
\end{thebibliography}
\end{document}